\documentclass[a4paper,11pt]{article}
\usepackage{pos}

\title{Momentum distribution of charm hadrons in a fluid-dynamic approach}

\author*[a]{Federica Capellino}
\author[b]{Andrea Dubla}
\author[c]{Stefan Floerchinger}
\author[d]{Eduardo Grossi}
\author[e]{Andreas Kirchner}
\author[a,b]{Silvia Masciocchi}

\affiliation[a]{Physikalisches Institut, Universit\"at Heidelberg\\
  Im Neuenheimer Feld 226, Heidelberg, Germany}

\affiliation[b]{GSI Helmoltzzentrum f\"ur Schwerionenforschung, \\
Planckstrasse 1, Darmstadt, Germany}
\affiliation[c]{Theoretisch-Physikalisches Institut, Friedrich Schiller Universit\"at Jena,\\
Fröbelstieg 1, Jena, Germany}
\affiliation[d]{Dipartimento di Fisica e Astronomia, Università degli Studi di Firenze,\\
Via Giovanni Sansone 1, Firenze, Italy}
\affiliation[e]{Institut f\"ur Theoretische Physik, Universit\"at Heidelberg,\\
Philosophenweg 16, Heidelberg, Germany}

\emailAdd{f.capellino@gsi.de}

\abstract{Exploiting a mapping between transport theory and fluid dynamics, we show how a fluid-dynamic description of the diffusion of charm quarks in the QCD plasma is feasible. 
We show results for spectra of charmed hadrons obtained with a fluid-dynamic description of the quark-gluon plasma (QGP) coupled with the conservation of a heavy-quark - antiquark current. We compare our calculations with the most recent experimental data in order to provide further constraints on the transport coefficients of the QGP.
}

\FullConference{HardProbes2023\\
 26-31 March 2023 \\
 Aschaffenburg, Germany\\}

\begin{document}
\maketitle

\section{Introduction}

Charm and beauty quarks produced in heavy-ion collisions are found to be suitable probes to study the features of the quark-gluon plasma (QGP). 
In fact, due to their large mass ($M_c\sim 1.5$ GeV, $M_b\sim 4.8$ GeV), they are produced via hard scatterings occurring at the very beginning of the collision, even before the QGP itself is created. 
Their dynamics in the QGP is regulated by transport coefficients which encode the microscopic description of the heavy quark-medium interaction. 
In the low transverse momentum region, heavy quarks provide a window to study equilibration processes. 
The idea of equilibration (or thermalization) is that -- if particles have enough time to interact with each other -- they will eventually relax to (at least local) thermal equilibrium. On the one hand, thermal equilibrium involves (local) chemical equilibrium. The latter implies that the particle abundance can be described by a distribution parametrized by a unique (local) chemical potential $\mu(x)$. In the case of heavy quarks, one knows from perturbative calculations (see Ref. \cite{FONLL} ) that they are produced very far from chemical equilibrium. Since their number density is much smaller with respect to the one of the light degrees of freedom, they will very likely remain out of chemical equilibrium for the full lifetime of the fireball. On the other hand, thermal equilibrium can also refer to (local) kinetic equilibrium. Kinetic equilibrium is achieved if the momentum distribution of the particle can be described by a Boltzmann (in the classic case) 
momentum distribution at the same (local) temperature $T(x)$ of the surrounding medium. Although this condition is not fulfilled at the time of the production of the heavy quark-antiquark ($Q\overline{Q}$) pairs, there are strong hints that it will by the end of the fireball evolution. 
The most recent measurements of elliptic flow of $\rm D$ mesons and ${\rm J/\psi}$ as a function of transverse momentum \cite{ALICE:2020iug} show a positive signal, in line with the one observed for light hadrons. 
This suggests that the \textit{hydrodynamization} time of charm quarks is small enough for them to get dragged along with the QGP. A similar conclusion can be drawn from the most recent Lattice-QCD calculations of the heavy-quark diffusion coefficient $D_s$ \cite{Altenkort:2023oms}. 
Motivated by these exciting findings, we propose to tackle the heavy-quark in-medium dynamics and their thermalization with an approach based on fluid-dynamics.

\section{Fluid-dynamic equations}

\begin{figure}[hbtp]

\centering
\includegraphics[width=0.48\textwidth]{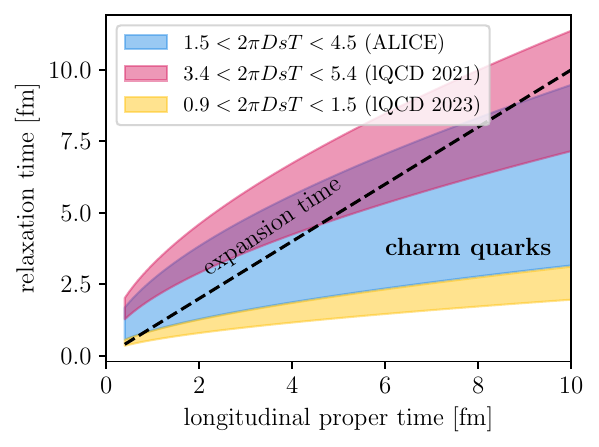}
\includegraphics[width=0.48\textwidth]{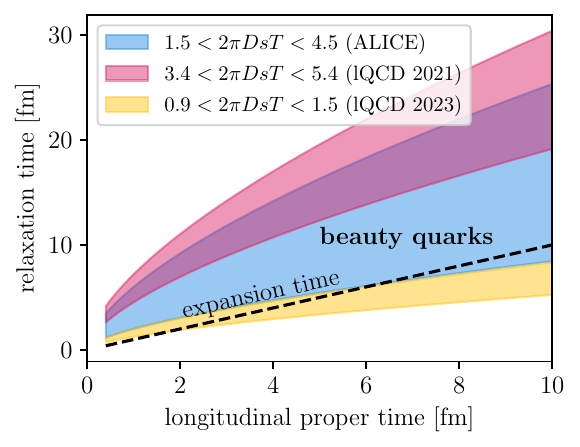}
\caption{Relaxation time of charm quarks (left panel) and beauty quarks (right panel) as a function of proper time in comparison with the typical expansion time of the fluid under the assumption of Bjorken flow. Different colored bands correspond to different values of the spatial diffusion coefficient $D_s$.}
\label{fig:reltimes}
\end{figure}

The hydrodynamic evolution of the QGP can be studied by implementing the equations for the conservation of the stress-energy tensor,
\begin{equation}
\nabla_\mu T^{\mu\nu} = 0\,.
\end{equation}
In the second-order hydrodynamic formalism, equations of motion for the dissipative currents, namely the bulk viscous pressure $\Pi$ and the shear-stress tensor $\pi^{\mu\nu}$, have to be considered to ensure causal behaviour. An Equation of State must be provided to close the system of equations. In this work, we extend this description to take into account an additional conserved current associated to the heavy-quark number. In fact, even though the conservation of $Q\overline{Q}$ pairs is not an \textit{exact} symmetry of QCD, it is still an \textit{accidental} symmetry. Due to the heavy quark large mass, the thermal production of a $Q\overline{Q}$ pair is negligible for the temperatures achieved during the fluid-dynamic evolution of the QGP. Furthermore, the annihilation rate of the pairs is negligible within the typical lifetime of the plasma. Hence, the number of $Q\overline{Q}$ pairs is considered as a conserved charge, and 
\begin{equation}
N^\mu = n u^\mu + \nu^\mu
\end{equation}
is the associated conserved current. It contains a term proportional to the fluid four-velocity $u^\mu$ via the $Q\overline{Q}$-pair density $n$ and a diffusion term $\nu^\mu$ orthogonal to the fluid-velocity. Beside the conservation law, 
\begin{equation}
\nabla_\mu N^\mu = 0\,,
\end{equation}
an equation of motion for the diffusion current is needed. In a linearized approach, this will be a relaxation-type equation,
\begin{equation}
\label{eqn:eom_partdiffcurr}
     \tau_n \Delta^\mu_{\,\rho} u^\sigma \partial_\sigma  \nu^\rho + \nu^{\mu}= \kappa_n  \nabla^\mu \left( \frac{\mu}{T} \right)\,,
\end{equation}
where $\Delta^{\mu \nu}=g^{\mu \nu}-u^\mu u^\nu$ is the projector onto the space orthogonal to the fluid velocity and we defined the transverse gradient $\nabla^\mu\equiv\Delta^{\mu\nu}\partial_\nu$. As one can observe, the evolution of the diffusion current is driven by gradients of the chemical potential associated to the number of $Q\overline{Q}$ pairs. 
The relaxation time $\tau_n$ plays the very role of a \textit{hydrodynamization} time, that is, a typical time scale after which the non-hydrodynamic modes vanish. 
The expressions for the relaxation time $\tau_n$ and the diffusion coefficient $\kappa_n$ in this hydrodynamic setup were derived in our previous work \cite{Capellino:2022nvf}.
There, an explicit relation between the relaxation time and the spatial diffusion coefficient $D_s$ was found. This relation allows to study, in a dynamical fashion, if the relaxation time stays smaller than the typical expansion time of the QGP. This is a necessary condition for the applicability of the fluid-dynamic framework. In Fig. \ref{fig:reltimes}, the comparison between $\tau_n$ and the typical expansion time of the fluid is shown as a function of proper time $\tau$ under the assumption of Bjorken flow \cite{Bjorken:1982qr}. Different colored bands correspond to different input values for $D_s$ given, respectively, by Lattice-QCD calculations both in quenched and non-quenched approximation \cite{Altenkort:2023oms,Altenkort:2020fgs} and from fits of multiple transport models to experimental measurements by the ALICE Collaboration \cite{ALICE:2021rxa}. We approximate the value of $2\pi D_s T$ to be valid at any temperature. The left panel shows result for charm quarks, whereas the right panel shows results for beauty quarks. As one can see, the relaxation time of charm quarks becomes much smaller than the typical expansion time of the fluid very early during the fluid evolution for a broad range of values of $D_s$. This suggests that the behaviour of charm quarks can be meaningfully described by fluid dynamics. For what regards beauty quarks, due to their larger mass, the range of applicability of hydrodynamics seems to become much more limited. However, the most recent Lattice-QCD calculations suggest that a (partial) hydrodynamization of beauty quarks could be possible, at least in the later stages of the fireball evolution.
As for the present work, the focus will be on the study of a hydrodynamic formalism for charm quarks only \cite{Capellino:2023cxe}.   

\section{Charmed-hadrons momentum distributions}
\begin{figure}[hbtp]
\centering

\includegraphics[width=0.48\textwidth]{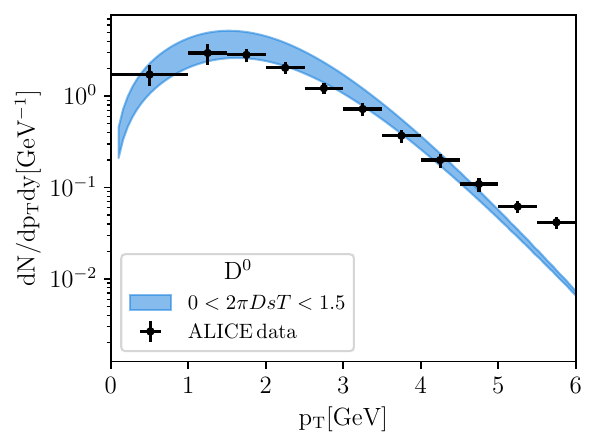}
\includegraphics[width=0.48\textwidth]{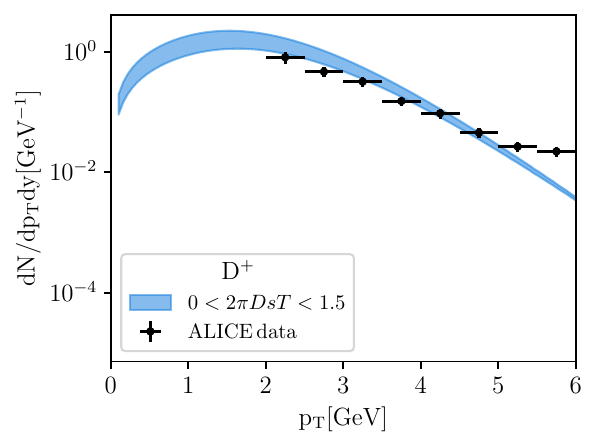}
\includegraphics[width=0.48\textwidth]{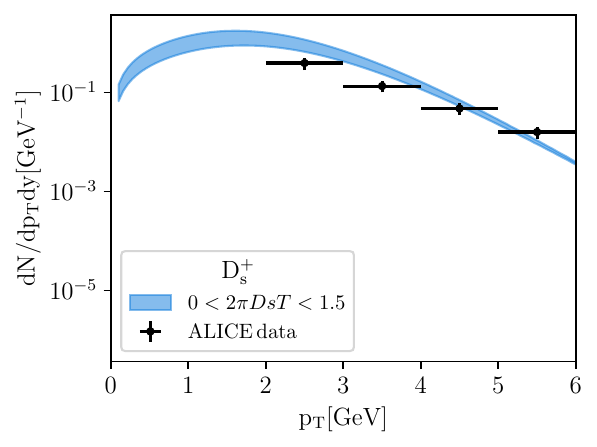}
\includegraphics[width=0.48\textwidth]{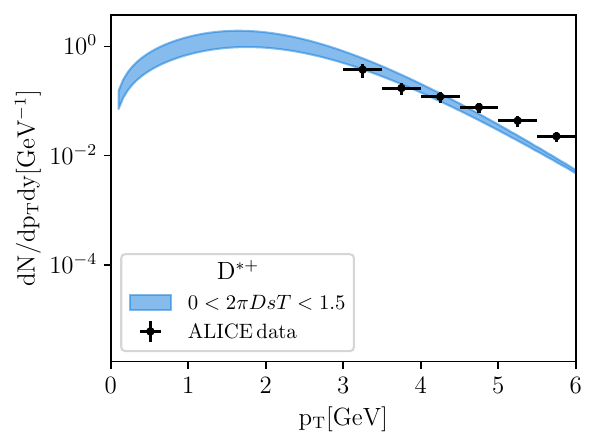}
\includegraphics[width=0.48\textwidth]{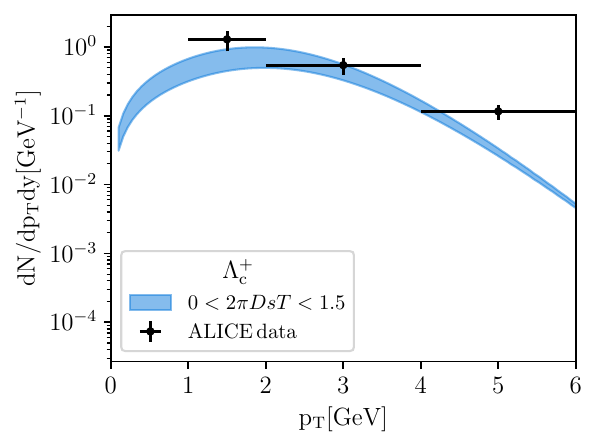}
\includegraphics[width=0.48\textwidth]{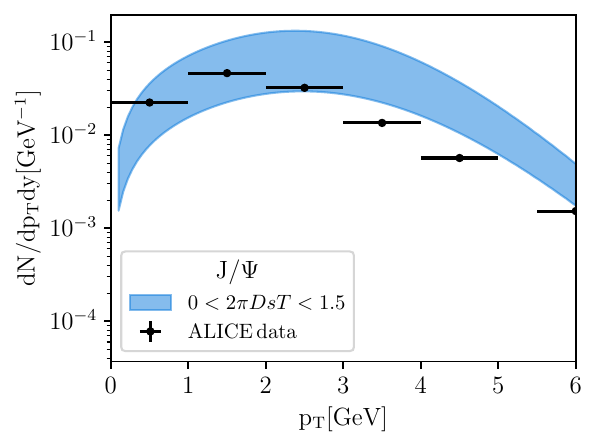}

\caption{Results for the momentum distributions of $\mathrm{D^0}$, $\mathrm{D^+}$, $\mathrm{D_s^+}$, $\mathrm{\Lambda_c^+}$ and $\mathrm{J/\psi}$ are shown in comparison with experimental measurements from the ALICE Collaboration \cite{ALICE:2021rxa,ALICE:2023gco,ALICE:2021bib,ALICE:2021kfc}. The color bands correspond to a spread on the input value of the spatial diffusion coefficient $D_s$ going from a non-diffusive case ($D_s = 0$) to the upper limit of the Lattice-QCD calculations ($2\pi D_s T_c = 1.5$).}
\label{fig:spectra}
\end{figure}

The equations of motion for the stress-energy tensor, charm current and associated dissipative quantities are solved numerically. Since the charm quarks number density is small compared to the one of the light degrees of freedom, their contribution to the total energy of the system is neglected. Therefore, the equations for $\mu$ and $\nu^\mu$ are solved assuming the equations for the fluid background fields ($T$, $u^\mu$, $\Pi$, $\pi^{\mu\nu}$) to be solved on-shell. This is equal to neglecting the back-reaction of the charm variables on the other fluid fields. The Equation of State and transport coefficients of the QGP are taken from \cite{Floerchinger:2018pje}, while the charm density and transport coefficients are taken as in \cite{Capellino:2023cxe}.
The initial conditions for the temperature fields are taken using $\mathrm{T_RENTo}$ \cite{Moreland:2014oya} to estimate the initial entropy density deposition in Pb-Pb collisions at 5.02 TeV in the 0-10$\%$ centrality class. The initial distribution of charm quarks is taken to be scaling with the number of binary collisions $n_{\rm coll}$. 
The momentum distributions are obtained employing a Cooper-Frye prescription at a freeze-out temperature of $156.5$ MeV \cite{Andronic:2017pug}. The resonance decays contributions are computed with the FastReso algorithm \cite{Mazeliauskas:2018irt}. The out-of-equilibrium corrections on the freeze-out surface are here not included, since they should be consistently derived for a multi-species fluid.\\
In Fig. \ref{fig:spectra} our calculations for the spectra of $\mathrm{D^0}$, $\mathrm{D^+}$, $\mathrm{D_s^+}$, $\mathrm{\Lambda_c^+}$ and $\mathrm{J/\psi}$ are shown in comparison with experimental measurements from the ALICE Collaboration \cite{ALICE:2021rxa,ALICE:2023gco,ALICE:2021bib,ALICE:2021kfc}. 
The colored bands correspond to a spread of the input value of the spatial diffusion coefficient $D_s$ going from a non-diffusive case ($D_s = 0$) to the upper limit of the Lattice-QCD calculations ($2\pi D_s T = 1.5$). The fluid-dynamic description seems to capture the physics of $\rm D$ mesons up to $p_{\rm T}\sim 4-5$ GeV. For what regards the $\mathrm{\Lambda_c^+}$, an overall underestimation of the integrated yield is observed, possibly indicating the existence of not-yet-measured resonance states \cite{He:2019tik,He:2019vgs}. The $\mathrm{J/\psi}$ momentum distribution, on the other hand, shows a peak for higher $p_{\rm T}$ values with respect to the measured one. 
\section{Conclusions and outlook}
This work has shown that a fluid-dynamic description for charm quarks is feasible. Remarkably, the momentum distributions of various charmed hadrons are found to be in agreement with the experimental data in a transverse momentum range up to $4-5$ GeV. 
Moreover, a consistent way of including the out-of-equilibrium correction at the freeze-out surface has to be developed, as well as possible relevant non-linear contributions in the equations of motion of the dissipative currents. Eventually, to validate the hypothesis of (full) charm thermalization, flow coefficients will be computed and systematically studied against experimental measurements in a continuation of this work.  

\section*{Acknowledgements}

The authors wish to thank A. Mazeliauskas and A. Andronic for the useful discussions. This work is funded via the DFG ISOQUANT Collaborative Research Center (SFB 1225). A.D. is partially supported by the Netherlands Organisation for Scientific Research (NWO) under the grant 19DRDN011, VI.Veni.192.039.

\end{document}